\documentclass[showpacs,12pt,nofootinbib,]{revtex4-1}
\usepackage{amsmath,amssymb}
\usepackage[dvips]{graphicx}
\usepackage[cp1251]{inputenc}
\usepackage{subfigure}

\newcommand\ba{\begin{eqnarray}}
\newcommand\ea{\end{eqnarray}}

\input{tcilatex}

\begin{document}

\title{Pion transition form factor in the constituent quark model}
\author{A.~E.~Dorokhov}
\email{dorokhov@theor.jinr.ru}
\author{E.~A.~Kuraev}
\email{kuraev@theor.jinr.ru}
\affiliation{JINR-BLTP, 141980 Dubna, Moscow region, Russian Federation}

\begin{abstract}
We calculate the transition form factor of the neutral pion where one photon
is virtual and another photon is real in the model where the light
constituent quark mass and the quark-pion vertex are taken to be momentum
independent. Radiative corrections to the lowest order triangle quark
Feynman amplitude are calculated. The resummation of the lowest radiative
corrections to the virtual photon vertex is done by applying the Sudakov
exponential hypothesis. Using fitting parameters, the quark mass and the
strong coupling constant, the results on the pion transition form factor are
compared with existing data published by CELLO, CLEO, BaBar and Belle
collaborations.
\end{abstract}

\pacs{13.60.-r, 13.66.Bc, 12.38.Lg, 12.38.Bx, 13.40.Gp}
\maketitle

\section{Introduction}

\begin{figure}[h]
\scalebox{0.7}{\includegraphics{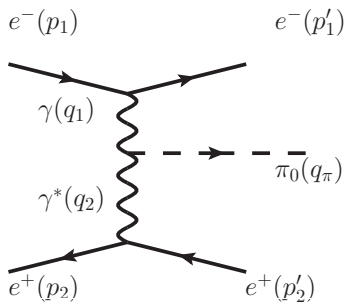}}
\caption{Neutral pion production in the $e^{+}e^{-}$ scattering.}
\label{Fg:ee scattering}
\end{figure}

A lot of attention has been paid to the problem of describing the transition
form factor of neutral pion. To get new information about the wave function
of the neutral pion \cite{Chernyak:1977as,Lepage:1979zb,Efremov:1979qk},
namely, the distribution of the neutral pion light-cone momentum fractions
between light $u$ and $d$ quarks, is the motivation for numerous theoretical
approaches to describe the transition form factor. Experimental information
about the form factor is obtained in the process $e^{+}e^{-}\rightarrow
e^{+}e^{-}\pi _{0}$ (Fig. \ref{Fg:ee scattering}). The kinematics, when one
photon is almost real and the other is highly virtual with space-like
momentum transfer squared $Q^{2},$%
\begin{equation}
|q_{1}^{2}|\approx 0\ll -q_{2}^{2}=Q^{2},  \label{Kinemat}
\end{equation}%
was measured by several experimental Collaborations: CELLO \cite%
{Behrend:1990sr}, CLEO \cite{Gronberg:1997fj} at low and intermediate $Q^{2}$
and more recently by BaBar \cite{:2009mc} and Belle \cite{Uehara:2012ag} at
higher $Q^{2}$.

The BaBar\ Collaboration fitted their experimental results for the form
factor multiplied by $Q^{2}$ as an increasing function of momentum transfer
squared (Fig. \ref{fig:FitAC})%
\begin{equation}
Q^{2}F_{\pi \gamma }^{A}\left( Q^{2}\right) =A\left( \frac{Q^{2}}{10~\mathrm{%
GeV}^{2}}\right) ^{\beta },  \label{ParamA}
\end{equation}%
with
\begin{eqnarray}
&&A_{\mathrm{BaBar}}=0.182~\mathrm{GeV,\quad \beta }_{\mathrm{BaBar}%
}=0.251,\quad \chi _{\mathrm{BaBar}}^{2}/[15]=1.04,  \label{BaBarparam} \\
&&A_{\mathrm{BaBar+}}=0.182~\mathrm{GeV,\quad \beta }_{\mathrm{BaBar+}%
}=0.252,\quad \chi _{\mathrm{BaBar+}}^{2}/[35]=0.87,  \label{BaBarchi}
\end{eqnarray}%
where in square brackets we point out the number of degrees of freedom. In
the first line only BaBar\ data are taken into account, while in the second
line the full set from BaBar, CLEO and CELLO\ data are taken into account.

The Belle\ Collaboration considered two parametrizations providing very
similar goodness of the fit. One is the same as (\ref{ParamA}) (Fig. \ref%
{fig:FitBC}) and the another corresponds to constant asymptotic behavior%
\begin{equation}
Q^{2}F_{\pi \gamma }^{B}\left( Q^{2}\right) =\frac{B}{1+\frac{C}{Q^{2}}}.
\label{ParamB}
\end{equation}%
The corresponding parameters are%
\begin{eqnarray}
&&A_{\mathrm{Belle}}=0.169~\mathrm{GeV,\quad \beta }_{\mathrm{Belle}%
}=0.18,\quad \chi _{\mathrm{Belle}}^{2}/[13]=0.429,  \label{Belleparam1} \\
&&A_{\mathrm{Belle+}}=0.172~\mathrm{GeV,\quad \beta }_{\mathrm{Belle+}%
}=0.221,\quad \chi _{\mathrm{Belle+}}^{2}/[33]=0.637,  \label{BellechiA} \\
&&B_{\mathrm{Belle}}=0.209~\mathrm{GeV,\quad C}_{\mathrm{Belle}}=2.2~\mathrm{%
GeV}^{2},\quad \chi _{\mathrm{Belle}}^{2}/[13]=0.435,  \label{Belleparam2} \\
&&B_{\mathrm{Belle+}}=0.186~\mathrm{GeV,\quad C}_{\mathrm{Belle+}}=0.948~%
\mathrm{GeV}^{2},\quad \chi _{\mathrm{Belle+}}^{2}/[33]=0.733.
\label{BellechiB}
\end{eqnarray}

The data on the pion-photon transition form factor obtained by CELLO, CLEO,
BaBar and Belle collaborations attract a lot attention \cite%
{Mikhailov:2009kf,Brodsky:2011yv,Arriola:2010aq,Noguera:2010fe,Agaev:2010aq,Kroll:2010bf,Klopot:2012hd}
with aim to extract the pion distribution amplitude, a nonperturbative
quantity important in description of hard exclusive hadronic processes.

The growing behavior of the form factor (\ref{ParamA}) is in clear
contradiction with the prediction of the approach based on the factorization
theorem applied to this process (see \cite{Mikhailov:2009kf} and references
therein). \ On the other hand, this behavior might indicate a
logarithmically enhanced asymptotic behavior of the form factor, as has been
argued in \cite%
{Dorokhov:2009dg,Radyushkin:2009zg,Polyakov:2009je,Bystritskiy:2009bk,Dorokhov:2010zzb}%
.

In \cite{Dorokhov:2009dg}, the neutral pion transition form factor was
considered in the model with momentum independent light quark mass and
quark-pion vertex. In this model, the leading order contribution to the form
factor is given by the triangle diagram of Fig. \ref{Fig:Born}. Its
asymptotic behavior for the kinematics (\ref{Kinemat}) is double logarithmic
one \cite{Dorokhov:2009dg,Bystritskiy:2009bk}, $\ln ^{2}\left(
Q^{2}/M_{q}^{2}\right) $, where $M_{q}$ is the quark mass serving as an
infrared cutoff parameter. In order to fit the BaBar data one needs to tune
the value of the mass parameter around $M_{q}\approx 135$ MeV (see below).
In \cite{Dorokhov:2010bz,Dorokhov:2010zzb,Dorokhov:2011dp}, this model has
been generalized considering momentum dependent quark mass and quark-pion
vertex. In \cite{Bystritskiy:2009bk}, the neutral pion transition form
factor has been considered in the leading - double-logarithmic approximation
in the scattering and annihilation channels.

In the present work, we are going to modify this model considering the
lowest order gluon radiative corrections. We shall use the well known
expression for the virtual photon-quark vertex (the so-called Sudakov form
factor \cite{Sudakov:1954sw}) which enters the triangle Feynman diagram,
describing the conversion of two photons to the neutral pseudoscalar meson.
The motivation of this and other similar studies is, first, to describe
existing data in full interval of $Q^{2}$, and, second, to understand if
there are any inconsistency between BaBar\ and Belle\ data at large $%
Q^{2}\geq 10$ GeV$^{2}.$

In Section \ref{Scattering} both channels of pseudoscalar mesons production
in electron-proton $e^{-}p\rightarrow e^{-}p\pi _{0}$ and electron-positron $%
e^{+}e^{-}\rightarrow \pi _{0}e^{+}e^{-}$ collisions are considered. The
second process could be the subject of future experimental investigation. In
Sections \ref{Born approximation}, \ref{NLO} and \ref{SudakovExp} we discuss
the leading order, next-to-leading order, and Sudakov exponentiation
calculations to the amplitude for the pion-photon transition. In Section \ref%
{Fits} we compare our model calculations with existing experimental data.

\section{Scattering channel}

\label{Scattering}

The matrix element for the neutral pion production in the high energy
electron-proton scattering (replacing the $e^{+}$ line in Fig. \ref{Fg:ee
scattering} by a proton $p$ line)
\begin{equation*}
e^{-}(p_{1})+p(p_{2})\rightarrow e^{-}(p_{1}^{\prime })+p(p_{2}^{\prime
})+\pi _{0}(p_{\pi }),
\end{equation*}%
can be written as
\begin{equation}
M^{ep\rightarrow ep\pi _{0}}=\frac{F(Q^{2})}{q_{2}^{2}q_{1}^{2}}\frac{%
8\alpha ^{2}}{f_{\pi }}N_{e}J_{\nu }(q_{2})\epsilon _{\mu \nu \lambda \sigma
}p_{1}^{\mu }q_{2}^{\lambda }q_{1}^{\sigma },  \label{MatEl}
\end{equation}%
for the kinematics of almost forward electron scattering
\begin{equation}
|q_{1}^{2}|=|\left( p_{1}-p_{1}^{\prime }\right) ^{2}|\ll
Q^{2}=-q_{2}^{2}=-\left( p_{2}-p_{2}^{\prime }\right) ^{2}\sim
s=(p_{1}+p_{2})^{2}.  \label{HEPlimit}
\end{equation}%
In (\ref{MatEl}), $f_{\pi }=92.2$ MeV is the pion decay constant, $%
N_{e}=(1/s)\bar{u}(p_{1}^{\prime })\hat{p}_{2}u(p_{1})$ and $F(Q^{2})$ is
the pion transition form factor.

\begin{figure}[h]
\scalebox{0.7}{\includegraphics{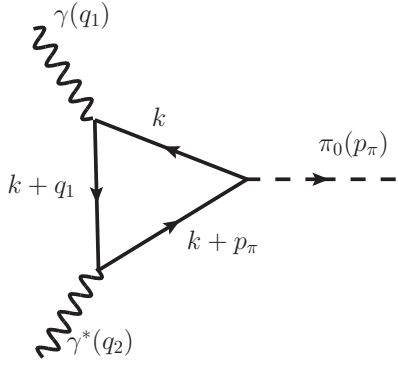}}
\caption{Lowest order QCD amplitude - the triangle vertex for $\protect%
\gamma \protect\gamma ^{\ast }\rightarrow \protect\pi _{0}$ process.}
\label{Fig:Born}
\end{figure}

In lowest order the form factor is given by the triangle quark loop integral
(Fig. \ref{Fig:Born})
\begin{eqnarray}
F_{0}(Q^{2}) &=&\int \frac{d^{4}k}{i\pi ^{2}}\frac{-M_{q}^{2}}{%
(k^{2}-M_{q}^{2})((k+q_{1})^{2}-M_{q}^{2})((k-q_{2})^{2}-M_{q}^{2})},
\label{BornTriangle} \\
F_{0}(0) &=&-2\frac{M_{q}^{2}}{m_{\pi }^{2}}\int\limits_{0}^{1}\frac{dx}{x}%
\ln (1-x(1-x)\frac{m_{\pi }^{2}}{M_{q}^{2}}),\qquad F_{0}(0)\overset{m_{\pi
}\rightarrow 0}{=}1,  \label{I0}
\end{eqnarray}%
where $m_{\pi }$ is the pion mass, $M_{q}$ is the light quark mass parameter
and $J_{\nu }(q)$ is the current corresponding to the proton vertex
\begin{equation}
J_{\nu }(q)=\bar{u}(p_{2}^{\prime })\left[ 2M_{P}\left(
F_{e}(q^{2})-F_{m}(q^{2})\right) \frac{1}{4M_{P}^{2}-q^{2}}%
(p_{2}+p_{2}^{\prime })_{\nu }+F_{m}(q^{2})\gamma _{\nu }\right] u(p_{2}),
\label{ProtonCurr}
\end{equation}%
where $M_{P}$ is the proton mass, $F_{e}$ and $F_{m}$ are the Sachs electric
and magnetic form factors of the proton.

Differential cross section for the process $ep\rightarrow ep\pi _{0}$ has
the form
\begin{equation}
d\sigma =\frac{\alpha ^{4}}{\pi ^{5}f_{\pi }^{2}}\frac{|F(Q^{2})|^{2}}{%
(q_{2}^{2}q_{1}^{2})^{2}}\Phi (Q^{2})\frac{1}{s}\frac{d^{3}p_{1}^{\prime }}{%
2E_{1}^{\prime }}\frac{d^{3}p_{2}^{\prime }}{2E_{2}^{\prime }}\frac{%
d^{3}p_{\pi }}{2E_{\pi }}\delta ^{4}(p_{1}+p_{2}-p_{1}^{\prime
}-p_{2}^{\prime }-p_{\pi })  \label{DifCross}
\end{equation}%
with
\begin{equation}
\Phi (Q^{2})=\left[ \left( \frac{1}{1+\frac{Q^{2}}{4M_{p}^{2}}}%
(F_{e}-F_{m})^{2}+F_{m}^{2}\right) (p_{2}+p_{2}^{\prime })_{\nu
}(p_{2}+p_{2}^{\prime })_{\nu _{1}}+F_{m}^{2}g_{\nu \nu _{1}}\right] (\nu
,p_{1},q,q_{1})(\nu _{1},p_{1},q,q_{1}),  \label{ProtBlock}
\end{equation}%
where $(\nu ,p_{1},q,q_{1})=\epsilon ^{\nu \alpha \beta \gamma }p_{1\alpha
}q_{\beta }q_{1\gamma }.$ In the limit of large $Q^{2}\sim s$ one has
\begin{equation}
\Phi (Q^{2}\sim s)=\frac{1}{4}(sQ^{2})^{2}\left( \frac{1}{1+\frac{Q^{2}}{%
4M_{p}^{2}}}\left( F_{e}(Q^{2})-F_{m}(Q^{2})\right)
^{2}+F_{m}^{2}(Q^{2})\right) .  \label{ProtBlockHEP}
\end{equation}%
In (\ref{DifCross}) we use the normalization of the matrix element for the
subprocess $\gamma ^{\ast }\gamma \rightarrow \pi ^{0}$ in accordance with
the current algebra for the case of two real photons ($q_{2}^{2}=q_{1}^{2}=0$%
) \cite{CA}
\begin{equation}
M^{\pi _{0}\rightarrow \gamma _{1}(q_{1}\epsilon _{1})\gamma
_{2}(q_{2}\epsilon _{2})}=i\frac{\alpha }{\pi f_{\pi }}(q_{1},q_{2},\epsilon
_{1},\epsilon _{2}),  \label{Anomaly}
\end{equation}%
corresponding to the width
\begin{equation}
\Gamma ^{\pi _{0}\rightarrow 2\gamma }=\frac{\alpha ^{2}m_{\pi }^{3}}{64\pi
^{3}f_{\pi }^{2}}\approx 7.1\quad \mathrm{KeV}.  \label{Width}
\end{equation}

For the $e^{\pm }e^{-}\rightarrow \pi _{0}e^{\pm }e^{-}$ collisions,
\begin{equation*}
e(p_{1}^{\pm })+e(p_{2})\rightarrow e(p_{1}^{\pm ^{\prime
}})+e(p_{2}^{\prime })+\pi _{0}(p_{\pi }),
\end{equation*}%
one should put $F_{e}(Q^{2})=F_{m}(Q^{2})=1$ in the above expressions for $%
\Phi $.

\section{Born approximation}

\label{Born approximation}

Let us now consider the amplitude $\gamma ^{\ast }\gamma \rightarrow \pi
^{0} $ in the context of the constituent quark model with momentum
independent quark mass $M_{q}$ \cite{Gerasimov:1978cp}. Within this model,
the pion form factor is given by the quark-loop (triangle) diagram (Fig. \ref%
{Fig:Born}). The result for the form factor in the considered asymmetric
kinematics $\left( q_{1}^{2}=0,q_{2}^{2}=-Q^{2}\right) $ is given by \cite%
{Ametller:1983ec}%
\begin{equation}
F_{0}\left( Q^{2}\right) =\frac{1}{Q^{2}}\frac{m_{\pi }^{2}}{1+\frac{m_{\pi
}^{2}}{Q^{2}}}\left[ \frac{1}{4\arcsin ^{2}\left( \frac{m_{\pi }}{2M_{q}}%
\right) }\ln ^{2}\frac{\beta _{q}+1}{\beta _{q}-1}+1\right]  \label{FbornA}
\end{equation}%
with $\beta _{q}=\sqrt{1+\frac{4M_{q}^{2}}{Q^{2}}}$ and the normalization is
$F_{0}\left( Q^{2}=0\right) =1.$

In \cite{Dorokhov:2009dg} this leading order (LO) expression was used to
explain growing-type form factor as it was measured by the BaBar
collaboration \cite{:2009mc}. The quark mass was used as the only fitting
parameter with the result (see below Table \ref{table:1})
\begin{equation}
M_{q}\approx 135\quad \mathrm{MeV.}  \label{MqFit}
\end{equation}

The expansion of the log term in the form factor (\ref{FbornA}) at large $%
Q^{2}>>M_{q}^{2}$ leads to%
\begin{equation}
F_{0}^{\mathrm{as}}\left( Q^{2}\right) =\frac{1}{Q^{2}}\frac{m_{\pi }^{2}}{%
2\arcsin ^{2}\left( \frac{m_{\pi }}{2M_{q}}\right) }\left\{ \frac{1}{2}%
L^{2}+2\arcsin ^{2}\left( \frac{m_{\pi }}{2M_{q}}\right) +O\left( \frac{%
M_{q}^{2}}{Q^{2}}\right) \right\} ,  \label{FbornAs}
\end{equation}%
where the large logarithm is%
\begin{equation*}
L=\ln \frac{Q^{2}}{M_{q}^{2}}.
\end{equation*}%
Numerically, $F_{0}^{\mathrm{as}}\left( Q^{2}\right) $ and $F_{0}\left(
Q^{2}\right) $ become indistinguishable at $Q^{2}>1$ GeV$^{2}$.

In the following, we would like to understand the role of radiative QCD
corrections to the $L^{2}$ term in the form factor which are more important
at large $Q^{2}$ than neglected in (\ref{FbornAs}) power corrections. To
this end, we first reproduce the leading $L^{2}$ asymptotic term in (\ref%
{FbornAs}) by using two different techniques which become useful when
considering the radiative corrections in the leading logarithmic
approximation.

One of these methods consists in joining the denominators of the integrand
in (\ref{BornTriangle}) by using Feynman parameters and performing the loop
momentum integration. In case of $q_{1}^{2}\rightarrow 0$ one obtains
\begin{eqnarray}
&&F_{0}(Q^{2})=2\int d^{3}x\delta \left( \sum_{i=1}^{3}x_{i}-1\right) \frac{%
M_{q}^{2}}{M_{q}^{2}-m_{\pi }^{2}x_{1}x_{3}+Q^{2}x_{2}x_{3}}=\frac{2M_{q}^{2}%
}{Q^{2}+m_{\pi }^{2}}\int\limits_{0}^{1}\frac{dx}{x}\ln \frac{1+\frac{Q^{2}}{%
M_{q}^{2}}x(1-x)}{1-\frac{m_{\pi }^{2}}{M_{q}^{2}}x(1-x)},  \notag \\
&=&\frac{M_{q}^{2}}{Q^{2}}L^{2}\left( 1+O\left( \frac{m_{\pi }^{2}}{M_{q}^{2}%
},\frac{M_{q}^{2}}{Q^{2}}\right) \right) .
\end{eqnarray}

An alternative calculation is based on the use of the Sudakov
parametrization of the loop momentum
\begin{equation*}
k=\alpha \tilde{q}_{1}+\beta \tilde{p}_{\pi }+k_{\bot },
\end{equation*}%
with $\tilde{q}_{1},\tilde{p}_{\pi }$ being the light-like four-vectors
constructed from the external momenta, $k_{\bot }\tilde{q}_{1}=k_{\bot }%
\tilde{p}_{\pi }=0$. By using the known relations
\begin{eqnarray}
&&d^{4}k=\frac{Q^{2}}{2}d\alpha d\beta d^{2}k_{\bot },\quad k_{\bot }^{2}=-%
\vec{k}^{2},  \notag \\
&&k^{2}-M_{q}^{2}+i0=Q^{2}\alpha \beta -\vec{k}^{2}-M_{q}^{2}+i0,\quad
(k+q_{1})^{2}-M_{q}^{2}=-Q^{2}\beta (1-\alpha )-\vec{k}^{2}-M_{q}^{2},
\notag \\
&&(k-q)^{2}-M_{q}^{2}=-Q^{2}\alpha (1-\beta )-\vec{k}^{2}-M_{q}^{2},
\end{eqnarray}%
and performing the integration in $\vec{k}^{2}$ by the relation
\begin{equation*}
\int \frac{d\vec{k}^{2}}{Q^{2}\alpha \beta -\vec{k}^{2}-M_{q}^{2}+i0}=-i\pi
\Theta \left( Q^{2}\alpha \beta -M_{q}^{2}\right) ,
\end{equation*}%
we obtain
\begin{equation*}
F_{0}^{\mathrm{as}}(Q^{2})=\frac{2M_{q}^{2}}{Q^{2}}\int%
\limits_{M_{q}^{2}/Q^{2}}^{1}\frac{d\alpha }{\alpha }\int%
\limits_{M_{q}^{2}/(Q^{2}\alpha )}^{1}\frac{d\beta }{\beta }=\frac{M_{q}^{2}%
}{Q^{2}}L^{2}.
\end{equation*}%
Here, we also take into account both possibilities of positive and negative
values of the Sudakov parameters $\alpha ,\beta $. Below, we shall use both
the Feynman and Sudakov approaches.
\begin{figure}[tbp]
\scalebox{0.5}{\includegraphics{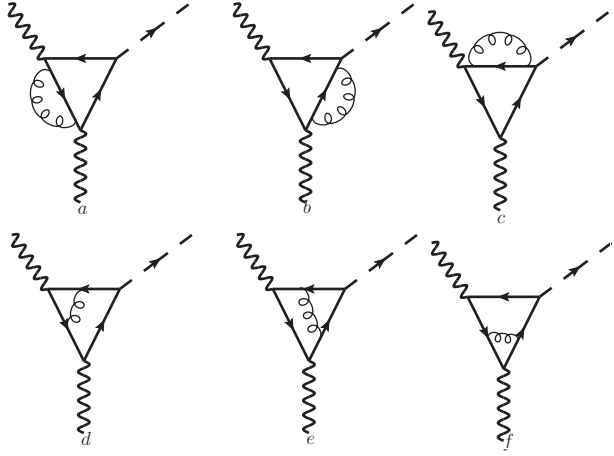}}
\caption{Next-to-leading QCD radiative corrections to the triangle amplitude
for $\protect\gamma \protect\gamma ^{\ast }\rightarrow \protect\pi _{0}$
process.}
\label{Fig: NLO}
\end{figure}

\section{Lowest order QCD radiative corrections}

\label{NLO}

The lowest order QCD corrections include three vertex subgraphs and three
quark self-energy subgraphs (Fig. \ref{Fig: NLO}). Kinematics of the main
contributions of the Feynman triangle amplitude correspond to the "almost
on-mass-shell" quark connecting the "almost on-mass-shell" photon and the
emission of a real pion, while the two other quark lines are essentially
off-mass-shell. Thus, one of the vertex function associated with the
off-mass-shell external photon underlies the Sudakov conditions: both quarks
are off-mass-shell. The two other vertices describe the situation when the
photon and one of quarks are almost on-mass-shell, while the other quark is
off-mass-shell. This configuration corresponds to the so-called Landau case
\cite{AhBer}. The contribution from the triangle amplitude with a mass
operator insertion to the "almost real" quark line does not contain
logarithmically enhanced terms.
\begin{figure}[tbp]
\scalebox{0.6}{\includegraphics{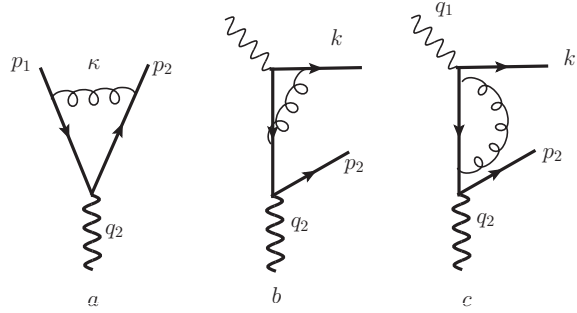}}
\caption{Lowest order QCD radiative correction to the subgraphs with
vertices and mass operators.}
\label{Fig:VertexNLO}
\end{figure}

Consider first the vertex subgraph with external highly virtual photon with
momentum $q_{2}$. The corresponding vertex function has the form (see Fig. %
\ref{Fig:VertexNLO}a)
\begin{eqnarray}
V_{\mu }(Q^{2}) &=&\frac{\alpha _{s}C_{F}}{4\pi }\int \frac{d^{4}\kappa }{%
i\pi ^{2}}\frac{N_{\mu }}{\kappa ^{2}\left( (p_{1}-\kappa
)^{2}-M_{q}^{2}\right) \left( (p_{2}-\kappa )^{2}-M_{q}^{2}\right) },
\label{VertA} \\
N_{\mu } &=&\bar{u}(p_{2})\gamma _{\lambda }(\hat{p}_{2}-\widehat{\kappa }%
+M_{q})\gamma _{\mu }(\hat{p}_{1}-\widehat{\kappa }+M_{q})\gamma ^{\lambda
}u(p_{1}),  \notag \\
Q^{2} &=&2p_{1}p_{2}>>|p_{1}^{2}|,|p_{2}^{2}|>>M_{q}^{2}.  \label{LimA}
\end{eqnarray}%
Here, $C_{F}=4/3$ is a Casimir invariant of the SU(3) color group.

The logarithmically enhanced contributions arise from two kinematically
different regions of virtual gluon momentum squared $|\kappa ^{2}|,$
corresponding to small and large values of $|\kappa ^{2}|$. Using the
Sudakov parametrization
\begin{equation*}
\kappa =\alpha _{1}p_{2}^{\prime }+\beta _{1}p_{1}^{\prime }+\overrightarrow{%
\kappa }_{\bot },\quad \kappa ^{2}=Q^{2}\alpha _{1}\beta _{1}-%
\overrightarrow{\kappa }^{2},\quad d^{4}\kappa =\pi \frac{Q^{2}}{2}d\alpha
_{1}d\beta _{1}d\overrightarrow{\kappa }^{2},
\end{equation*}%
with $p_{1}^{\prime },p_{2}^{\prime }$ light-like 4-vectors built from $%
p_{1},p_{2}$, we write $N_{\mu },$ in the limit (\ref{LimA}), as
\begin{equation*}
N_{\mu }=[2Q^{2}(1-\alpha _{1})(1-\beta _{1})+\kappa ^{2}]J_{\mu }
\end{equation*}%
with%
\begin{equation*}
J_{\mu }=\bar{u}(p_{2})\gamma _{\mu }u(p_{1}).
\end{equation*}

For the first term in the square brackets of $N_{\mu }$ we obtain (small $%
|\kappa ^{2}|$ region):
\begin{eqnarray}
V_{1\mu }(Q^{2};\alpha ,\beta ) &=&-2\frac{\alpha _{s}C_{F}}{4\pi }J_{\mu
}\int_{0}^{1}\frac{d\alpha _{1}(1-\alpha _{1})}{\alpha _{1}+\alpha }%
\int_{0}^{1}\frac{d\beta _{1}(1-\beta _{1})}{\beta _{1}+\beta }\theta
(Q^{2}\alpha _{1}\beta _{1}-M_{q}^{2})  \notag \\
&=&-\frac{\alpha _{s}C_{F}}{2\pi }\left[ \ln \alpha \ln \beta +\ln (\alpha
\beta )\right] J_{\mu },  \label{VertA1}
\end{eqnarray}%
where the factor 2 is due to the two regions of negative and positive values
of the Sudakov parameters. In the derivation of (\ref{VertA1}), we used the
relations $|p_{1}^{2}/Q^{2}|=|\alpha |,|p_{2}^{2}/Q^{2}|=|\beta |$, with $%
\alpha ,\beta $ being the Sudakov variables associated with the loop
momentum of the quark loop.

The contribution from the region with large $|\kappa ^{2}|$ comes from the
second term in the square brackets of $N_{\mu },$
\begin{equation*}
V_{2\mu }(Q^{2})=-\frac{\alpha _{s}C_{F}}{4\pi }J_{\mu }\int \frac{%
d^{4}\kappa }{i\pi ^{2}}\frac{1}{\left( (p_{1}-\kappa )^{2}-M_{q}^{2}\right)
\left( (p_{2}-\kappa )^{2}-M_{q}^{2}\right) }.
\end{equation*}%
Here, we have to introduce the ultraviolet cut-off parameter $\Delta $, such
as $|\kappa ^{2}|<\Lambda ^{2}$. The usual procedure of joining the
denominators and performing the integration leads to
\begin{equation}
V_{2\mu }(Q^{2})=-\frac{\alpha _{s}C_{F}}{4\pi }J_{\mu
}\int\limits_{0}^{1}dx\int \frac{d^{4}\kappa }{i\pi ^{2}}\frac{1}{[(\kappa
-p_{x})^{2}-Q^{2}x(1-x)]^{2}}=-\frac{\alpha _{s}C_{F}}{4\pi }J_{\mu }\ln
\frac{\Lambda ^{2}}{Q^{2}}.  \label{VertA2}
\end{equation}%
where $p_{x}=xp_{1}+(1-x)p_{2}.$ Note, that we systematically omit the
logarithmically suppressed terms. After regularization, the cut-off
parameter must be replaced by the quark mass.

The total answer for $V_{\mu }$ is
\begin{equation}
V_{\mu }(Q^{2};\alpha ,\beta )=-\frac{\alpha _{s}C_{F}}{2\pi }\left[ \ln
\alpha \ln \beta +\ln (\alpha \beta )+\frac{1}{2}L\right] J_{\mu }.
\label{Vlog}
\end{equation}

The contribution of other Feynman amplitudes (Fig. \ref{Fig:VertexNLO}b,c)
does not contain logarithmic enhancement. We will illustrate this statement
in the frame of QED \cite{Kuraev:1987cg,AhBer}. Let us consider the
contribution of two remaining diagrams with one vertex function (Fig. \ref%
{Fig:VertexNLO}b) and mass operator insertion (Fig. \ref{Fig:VertexNLO}c)
\begin{eqnarray}
&&V_{\mu ,\nu }=-4\pi \alpha _{s}C_{F}\bar{u}(p_{2})\gamma _{\mu }[\frac{1}{t%
}(\hat{p}_{2}-\hat{q}_{2}+M_{q})\hat{\Gamma}_{\nu }+\frac{\widehat{M}\left(
p_{2}-q_{2}\right) }{(\hat{p}_{2}-\hat{q}_{2}-M_{q})^{2}}\gamma _{\nu
}]u(k),\quad  \label{VertB} \\
&&t=(p_{2}-q_{2})^{2}-M_{q}^{2},\quad p_{2}-q_{2}=q_{1}-k.  \notag
\end{eqnarray}%
Using the explicit expressions for the vertex $\Gamma _{\nu }$ and mass
operator $\widehat{M}(p)$ given in the Appendix, we obtain
\begin{equation*}
V_{\mu \nu }=-\frac{2\alpha ^{2}i}{M_{q}}\bar{u}(p_{2})\gamma _{\mu }\left[
A_{1}(\gamma _{\nu }-\hat{q}_{1}\frac{1}{kq_{1}}k_{\nu })+A_{2}\frac{1}{M_{q}%
}\hat{q}_{1}\gamma _{\nu }\right] ,
\end{equation*}%
where
\begin{eqnarray*}
A_{1} &=&\frac{1}{2(\widetilde{t}+1)}\left[ 1-\frac{\widetilde{t}}{%
\widetilde{t}+1}l_{t}\right] ,\ \  \\
\widetilde{t} &=&\frac{t}{M_{q}^{2}},\ \ l_{t}=\ln \frac{-t}{M_{q}^{2}}.
\end{eqnarray*}%
We see, that the relevant contribution is suppressed by small factor $1/%
\widetilde{t}$. Other terms do not contribute to the amplitude. Similar
statement is valid for two remaining diagrams (see Fig. \ref{Fig: NLO} b,e).

Collecting all logarithmically enhanced contributions, we finally obtain
\begin{eqnarray}
&&\int\limits_{M_{q}^{2}/Q^{2}}^{1}\frac{d\alpha }{\alpha }%
\int\limits_{M_{q}^{2}/(Q^{2}\alpha )}^{1}\frac{d\beta }{\beta }\left[ 1-%
\frac{\alpha _{s}C_{F}}{2\pi }\left( \ln \alpha \ln \beta +\ln (\alpha \beta
)+\frac{1}{2}\ln \frac{Q^{2}}{M_{q}^{2}}\right) \right]  \label{FNLO} \\
&=&\frac{1}{2}L^{2}-\frac{\alpha _{s}C_{F}}{2\pi }\frac{1}{12}L^{3}\left(
\frac{1}{2}L-1\right) .  \notag
\end{eqnarray}%
Thus, the form factor at large $Q^{2}$ (\ref{FbornAs}), modified by
next-to-leading (NLO) radiative QCD corrections, becomes%
\begin{equation}
F_{1}^{\mathrm{as}}\left( Q^{2}\right) =\frac{1}{Q^{2}}\frac{m_{\pi }^{2}}{%
2\arcsin ^{2}\left( \frac{m_{\pi }}{2M_{q}}\right) }\left\{ \frac{1}{2}L^{2}-%
\frac{\alpha _{s}C_{F}}{2\pi }\frac{1}{12}L^{3}\left( \frac{1}{2}L-1\right)
+2\arcsin ^{2}\left( \frac{m_{\pi }}{2M_{q}}\right) \right\} .  \label{FNLOa}
\end{equation}

\section{Higher order QCD generalization}

\label{SudakovExp}

In order to estimate the effect of the higher orders of QCD perturbation
theory we apply to the next-to-leading logarithmic approximation (\ref{FNLO}%
) the Sudakov exponentiation hypothesis with the result
\begin{eqnarray}
&&\int\limits_{M_{q}^{2}/Q^{2}}^{1}\frac{d\alpha }{\alpha }%
\int\limits_{M_{q}^{2}/(Q^{2}\alpha )}^{1}\frac{d\beta }{\beta }\exp \left\{
\alpha _{s}\frac{C_{F}}{2\pi }\left[ -\ln \alpha \ln \beta -\ln (\alpha
\beta )-\frac{1}{2}L\right] \right\}  \notag \\
&=&\frac{1}{\rho }e^{-\frac{1}{2}L\rho }\int_{0}^{1}\frac{dx}{x-\frac{1}{L}}%
e^{xL\rho }\left( 1-e^{-L^{2}\rho (1-x)\left( x-\frac{1}{L}\right) }\right) ,
\label{FAllOrder}
\end{eqnarray}%
where $\rho =\frac{\alpha _{s}}{2\pi }C_{F}.$ For the form factor the
resummation effects leads to the generalization of (\ref{FNLOa}) as%
\begin{eqnarray}
&&F_{\mathrm{Exp}}^{\mathrm{as}}\left( Q^{2}\right) =\frac{1}{Q^{2}}\frac{%
m_{\pi }^{2}}{2\arcsin ^{2}\left( \frac{m_{\pi }}{2M_{q}}\right) }
\label{FSudOrder} \\
&&\cdot \left\{ \frac{1}{\rho }e^{-\frac{1}{2}L\rho }\int_{0}^{1}\frac{dx}{x-%
\frac{1}{L}}e^{xL\rho }\left( 1-e^{-L^{2}\rho (1-x)\left( x-\frac{1}{L}%
\right) }\right) +2\arcsin ^{2}\left( \frac{m_{\pi }}{2M_{q}}\right)
\right\} .  \notag
\end{eqnarray}

\section{The results and discussion}

\label{Fits}

\begin{figure}[tbp]
\scalebox{0.45}{\includegraphics{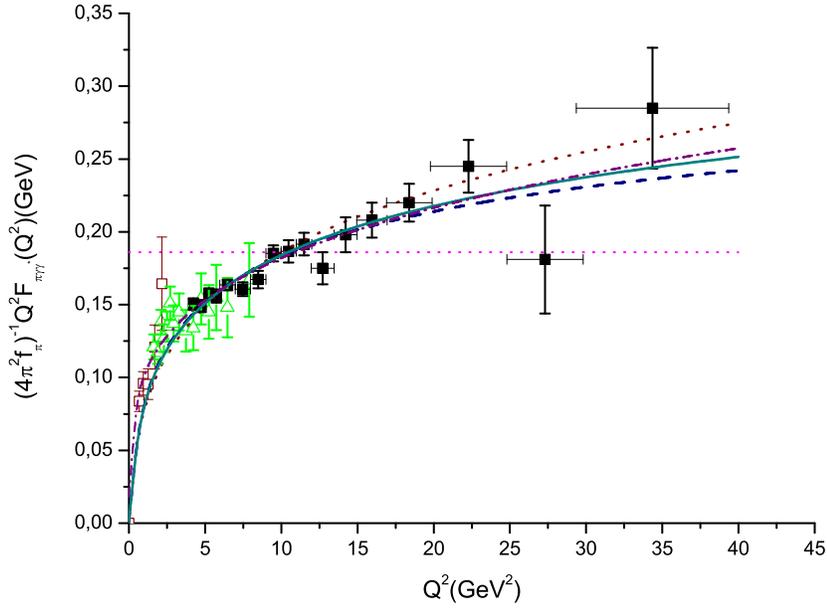}} 
\caption{ The fit of our model for the cases of the LO approximation (Eqs. (%
\protect\ref{FbornAs}) -- doted line), the NLO approximation (Eq. ( \protect
\ref{FNLOa}) -- dash line), the resummation approximation ((\protect\ref%
{FSudOrder}) -- solid line) for the $\protect\pi ^{0}$ form factor and its
comparison with the experimental data of CELLO \protect\cite{Behrend:1990sr}
(open boxes), CLEO \protect\cite{Gronberg:1997fj} (open triangles) and BaBar
\protect\cite{:2009mc} (filed boxes) Collaborations. The dash-dot line shows
the fit of the data by BaBar collaboration.}
\label{fig:FitAC}
\end{figure}

\begin{figure}[tbp]
\scalebox{0.45}{\includegraphics{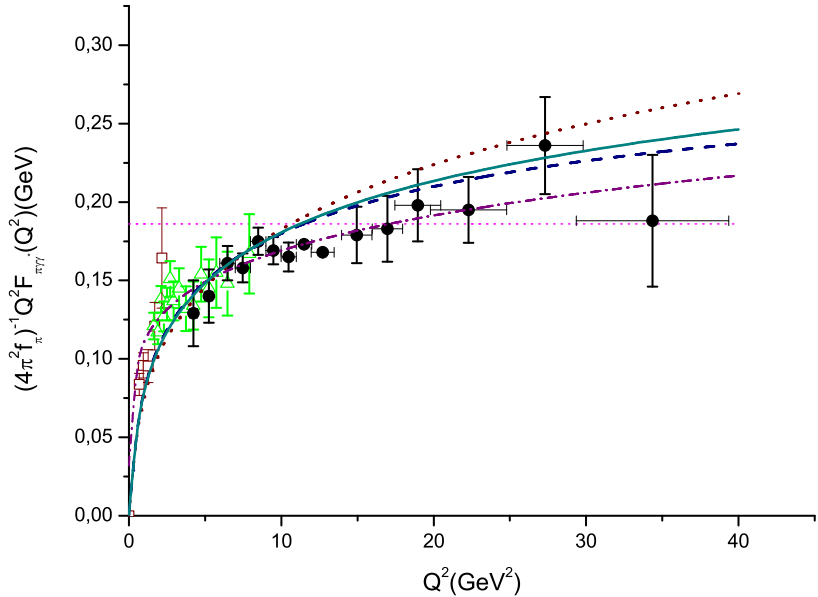}} 
\caption{ The fit of our model for the cases of the LO approximation (Eqs. (%
\protect\ref{FbornAs}) -- doted line), the NLO approximation (Eq. ( \protect
\ref{FNLOa}) -- dash line), the resummation approximation ((\protect\ref%
{FSudOrder}) -- solid line) for the $\protect\pi ^{0}$ form factor and its
comparison with the experimental data of CELLO \protect\cite{Behrend:1990sr}
(open boxes), CLEO \protect\cite{Gronberg:1997fj} (open triangles) and Belle
\protect\cite{Uehara:2012ag} (filed circles) Collaborations. The dash-dot
line shows the fit of the data by Belle collaboration.}
\label{fig:FitBC}
\end{figure}

There are two parameters within the parametrization of the pion transition
form factor suggested by the quark model with radiative corrections. They
are the quark mass $M_{q}$ and the strong coupling constant $\alpha _{s}$.
We try to fit the experimental data for the form factor by varying these
parameters. The results of the fit are presented in Tables \ref{table:1}-\ref%
{table:3} and Figs. \ref{fig:FitAC}-\ref{fig:FitBC}.

In fitting procedure we use the parametrizations (\ref{ParamA}) and (\ref%
{ParamB}) as reference parametrizations. From corresponding goodness of fit (%
\ref{BaBarparam}), (\ref{BaBarchi}) and (\ref{Belleparam1}), (\ref{BellechiA}%
) one concludes, first, that $\chi _{\mathrm{BaBar}}^{2}>\chi _{\mathrm{Belle%
}}^{2}$ and, second, that the data set extended by experimental points of
CELLO and CLEO $\chi _{\mathrm{BaBar}}^{2}>\chi _{\mathrm{BaBar+}}^{2}$ and $%
\chi _{\mathrm{Belle}}^{2}<\chi _{\mathrm{Belle+}}^{2}$. The first fact is
due to systematically lower error bars for the BaBar point set than for the
Belle. The second property we interpret as indication on better consistency
of the BaBar data with previous data at lower $Q^{2}$ than for the Belle data%
\footnote{%
At this point we disagree with conclusions made in \cite{Stefanis:2012yw}.}.

First of all, let us use our model to fit the BaBar and Belle data,
considering the quark mass as a free parameter and the strong coupling to be
fixed at $\alpha _{s}=0.35$, which corresponds to a renormalization scale of
about 1 GeV. The results are given in Table I. For the leading order (LO)
fit we use expression (\ref{FbornAs}), for the next-to-leading order (NLO)
fit we use expression (\ref{FNLOa}), and the Sudakov resummed expression
(Exp) is (\ref{FSudOrder}). In order to compare goodness of fit based on
data used from different collaborations we introduce the relative parameter%
\begin{equation}
\overline{\chi }^{2}=\chi ^{2}/\chi _{a}^{2},  \label{relative chi2}
\end{equation}%
with $a$ is for BaBar or Belle based data set. In (\ref{relative chi2}), $%
\chi ^{2}$ is for our model and for the power-like fits $\chi _{a}^{2}$ is
from (\ref{BaBarparam}), (\ref{BaBarchi}) and (\ref{Belleparam1}), (\ref%
{BellechiA}), correspondingly.

\begin{table*}[th]
\centering%
\begin{tabular}{|l|l|l|l|l|l|l|}
\hline
& \multicolumn{3}{|l|}{BaBar} & \multicolumn{3}{|l|}{Belle} \\ \hline
& $M_{q}$ & $\chi ^{2}/$[16] & $\overline{\chi }^{2}$ & $M_{q}$ & $\chi
^{2}/ $[14] & $\overline{\chi }^{2}$ \\ \hline
LO & 0.135 & 1.697 & 1.629 & 0.126 & 0.696 & 1.611 \\ \hline
NLO & 0.149 & 1.185 & 1.137 & 0.142 & 0.434 & 1.005 \\ \hline
Exp & 0.147 & 1.196 & 1.148 & 0.140 & 0.478 & 1.106 \\ \hline
\end{tabular}%
\caption{One-parameter fit of the BaBar and Belle data. In square brackets,
there is pointed out the number of degrees of freedom.}
\label{table:1}
\end{table*}

From Table \ref{table:1} one finds that the goodness of the fit becomes
better when going from the LO fit to the NLO fit, and almost does not change
after Sudakov resummation. This justifies our model with radiative
corrections. At the same time, the parameter $M_{q}$ becomes higher.

\begin{table*}[th]
\centering%
\begin{tabular}{|l|l|l|l|l|l|l|}
\hline
& \multicolumn{3}{|l|}{BaBar+} & \multicolumn{3}{|l|}{Belle+} \\ \hline
& $M_{q}$ & $\chi ^{2}/$[36] & $\overline{\chi }^{2}$ & $M_{q}$ & $\chi
^{2}/ $[34] & $\overline{\chi }^{2}$ \\ \hline
LO & 0.136 & 2.377 & 2.732 & 0.133 & 2.891 & 4.538 \\ \hline
NLO & 0.150 & 1.487 & 1.709 & 0.147 & 1.684 & 2.644 \\ \hline
Exp & 0.148 & 1.587 & 1.824 & 0.145 & 1.878 & 2.948 \\ \hline
\end{tabular}%
\caption{One-parameter fit of the BaBar and Belle data including also the
data from CELLO and CLEO collaborations. In square brackets, there is
pointed out the number of degrees of freedom.}
\label{table:2}
\end{table*}

In Table \ref{table:2} we made a fit of the BaBar and Belle data including
also the set of points from CELLO and CLEO collaborations. We see that,
qualitatively, for our model the situation does not change too much and the
value of $M_{q}$ is practically the same for both cases. The later fact is
rather important. It means that the fit procedure is basically related to
the data points at intermediate $Q^{2}$ in the region from 5 to 10 GeV$^{2}$
where the data are more precise and consistent for all collaborations. At
the same time, the region of higher $Q^{2}$ (15-40 GeV$^{2}$) is less
important for the fit. The form factor $F_{\pi \gamma \gamma ^{\ast
}}(Q^{2}) $ in accordance with different parameterizations given in Table %
\ref{table:2} is drawn in Figs.~\ref{fig:FitAC} and \ref{fig:FitBC}. In
these figures we also present a power-like fits to the BaBar and Belle data (%
\ref{BaBarparam}) and (\ref{Belleparam1}), correspondingly.

\begin{table*}[th]
\centering%
\begin{tabular}{|l|l|l|l|l|l|l|l|l|}
\hline
& \multicolumn{4}{|l|}{BaBar} & \multicolumn{4}{|l|}{Belle} \\ \hline
& $M_{q}$ & $\alpha _{s}$ & $\chi ^{2}/$[15] & $\overline{\chi }^{2}$ & $%
M_{q}$ & $\alpha _{s}$ & $\chi ^{2}/$[13] & $\overline{\chi }^{2}$ \\ \hline
NLO & 0.149 & 0.349 & 1.264 & 1.215 & 0.149 & 0.505 & 0.428 & 0.998 \\ \hline
Exp & 0.152 & 0.513 & 1.187 & 1.196 & 0.159 & 0.963 & 0.422 & 0.984 \\ \hline
\end{tabular}%
\caption{Two-parameter fit of the BaBar and Belle data. In square brackets,
there is pointed out the number of degrees of freedom.}
\label{table:3}
\end{table*}

In Table \ref{table:3} we made a two-parametric fit to the BaBar and Belle
data. We see that such a two-parametric fit has equal or even lower $\chi
^{2}$ with respect to corresponding numbers in Table \ref{table:1}. However,
the price for that is a growing of the parameter $\alpha _{s},$ especially
for the case of Belle data. We consider such parametrization as not very
physical.

\section{Conclusions}

\label{Cocl}

In the present work we calculated the transition form factor of the neutral
pion where one photon is virtual and another photon is real in the framework
of the model where the light constituent quark mass and the quark-pion
coupling are momentum independent\footnote{%
Similar model was considered in \cite%
{Pivovarov:2001mw,Boughezal:2011vw,Greynat:2012ww} in view of calculations
of hadronic corrections to the muon anomalous magnetic moment.}. We
generalize the previous leading order results \cite%
{Dorokhov:2009dg,Bystritskiy:2009bk} obtained by considering the triangle
diagram by including the radiative gluonic corrections to the first order in
perturbation theory\footnote{%
The problem of radiative corrections to the pion transition form factor was
considered sometime ago in \cite%
{delAguila:1981nk,Braaten:1982yp,Kadantseva:1985kb}, within factorization
approach with massless quarks. In this case the "large logarithms" $\log
Q^{2}/\mu ^{2}$ contain the QCD scale parameter $\mu $.}. The effect of
higher order radiative corrections to the virtual photon vertex is estimated
by applying the Sudakov exponentiation hypothesis. The results obtained are
compared with existing experimental data on the pion transition form factor
published by CELLO, CLEO, BaBar and Belle collaborations.

In general the model considered contains two parameters: the quark mass $%
M_{q}$ and the strong coupling constant $\alpha _{s}$.

First, we fit the data at fixed $\alpha _{s}$ varying only $M_{q}$. Taking
into account the radiative corrections increases a little the fitting
parameter up to $M_{q}\approx 150$ MeV and improves a goodness of the fit.
Considering separate fit of BaBar and Belle data (Table \ref{table:1}) one
gets very close $\overline{\chi }^{2}$ and the difference in $M_{q}$ is less
than 10 MeV. Including also lower momentum data from CELLO and CLEO (Table %
\ref{table:2}) leads to inequality $\overline{\chi }_{\mathrm{BaBar+}}^{2}<%
\overline{\chi }_{\mathrm{Belle+}}^{2}$ and almost coinciding $M_{q}$. To
our opinion it means that fitting procedure is most sensitive to the
intermediate momentum interval, where both set of data are in agreement. As
for parametrizations discussed in Introduction, BaBar data turns out to be
more in accordance with lower momentum data than the Belle data. Our model
is more close to the tendency of the BaBar data. Resummation effects do not
lead to significant changes in goodness of the fit comparing with the NLO
results.

Second, we try to fit data varying both model parameters (Table \ref{table:3}%
). The fit of data becomes better, especially in the region of large $Q^{2}$%
. However, $\alpha _{s}$ has tendency to be close to unit. This fact is
considered as not physically justified and thus not taken into account in
our final results.

Let us emphasize that at the moment there are two sets of experimental data
(BaBar and Belle) on the pion-photon transition form factor at high  $Q^{2}$%
. They are fully consistent in the range of momentum transfer squared $%
Q^{2}\sim \left[ 5-10\right] $ GeV$^{2}$, but have different tendency at
higher $Q^{2}$. Conditionally, the BaBar data show "growing" behavior at
large $Q^{2}$ (see (\ref{ParamA})), while the Belle data can be interpreted
in twofold way as "quasi-growing--quasi-constant" (see (\ref{ParamA}) and (%
\ref{ParamB})). From experimental point of view, the data in the range $%
Q^{2}\sim \left[ 15-40\right] $ GeV$^{2}$ are consistent at the level of $%
1\sigma $ standard deviation. However, from theoretical point of view there
are big debates on this difference \cite%
{Mikhailov:2009kf,Brodsky:2011yv,Arriola:2010aq,Noguera:2010fe,Agaev:2010aq,Kroll:2010bf,Klopot:2012hd,Dorokhov:2009dg,Radyushkin:2009zg,Polyakov:2009je,Bystritskiy:2009bk,Dorokhov:2010zzb}%
. It is clear that only new high statistic experiments can resolve this
problem.

The emergence of "growing" data was so unexpected that, at first glance, it
seemed that it is impossible to explain such behavior from the field
theoretical point of view. However, in \cite%
{Dorokhov:2009dg,Radyushkin:2009zg,Polyakov:2009je,Bystritskiy:2009bk,Dorokhov:2010zzb}
there was noted that, if such "growing" behavior exists, then it may be
related to unusual properties of the pion distribution amplitude in the
vicinity of its edge points. This behavior was conditionally called as
"flat". In this case the inverse moment of the pion distribution amplitude
is not well defined. In particular, such behavior may be modeled if to
assume that the pion is almost structureless. In \cite{Dorokhov:2009dg} (and
later in some other works) it was shown that BaBar data can be described in
the model with momentum independent quark mass and quark-pion coupling by
using only one parameter: the constituent quark mass $M_{q}$ if its value is
taken as  $M_{q}\approx 135$ MeV. This number was considered as a rather
small from phenomenological point of view. One of the main motivation of the
present work is to see how sensitive this parameter to gluonic radiative
corrections, when fit the BaBar and Belle data. The result is that the mass
parameter becomes a bit heavier $M_{q}\approx 150$ MeV and the quality of
the fit becomes better.

As it was shown in \cite{Dorokhov:2010zzb,Dorokhov:2010bz}, a more advanced
model, with momentum dependent quark mass and quark-pion vertex, has the
same qualitative features as the model considered in this work. In the model
\cite{Dorokhov:2010zzb,Dorokhov:2010bz}, the soften quark propagator and
quark-pion vertex lead to the single logarithmic asymptotic dependence on $%
Q^{2}$ instead of the double logarithmic behavior, if the quark-pion vertex
so that it corresponds to the "flat" pion distribution amplitude. The pion distribution amplitude 
even can vanish at the end-points, but still simulate the logarithmic growth in rather wide range of momentum transfer including 
quite large values of it. When fit
BaBar data, the mass parameter is still close to $135$ MeV. The calculations
given in the present paper can be extended by consideration not only
momentum dependent nonperturbative quark propagator and quark-pion vertex,
but also momentum dependent nonperturbative gluon propagator.

Finally note that there are available data obtained by the BaBar
collaboration for the $\eta $, $\eta ^{\prime }$ and $\eta _{c}$ transition
form factors. The comparison of our model calculations with data for these
mesons is given in \cite{Dorokhov:2011dp}. From our point of view the data
show tendency that with increase of the meson mass the form factor changes
its behavior from the "growing" regime to the "constant" regime. In the
framework of our model this change of behavior is related to strong
dependence of the shape of the meson distribution amplitude on the meson
mass. With growing meson mass the meson distribution amplitude changes its
shape from the "flat" one to the "$\delta $-function" shape for heavy mesons
\cite{Dorokhov:2011dp}.

The study of present work shows that inclusion of the QCD corrections are
essential in interpretation of experimental data on the pion transition form
factor.

\begin{acknowledgments}
We thank Yu.M. Bystritsky, V.V.~Bytev, N.I. Kochelev and Yu.S. Surovtcev for
their interest to the subject of present work and discussions.

This work is supported in part by the Russian Foundation for Basic Research
(project No. 11-02-00112).
\end{acknowledgments}

\appendix

\section{Appendix}

The explicit expressions for the vertex function in the Landau kinematics
and the mass operator are \cite{AhBer,Kuraev:1987cg}
\begin{eqnarray}
\hat{\Gamma}_{\nu } &=&\frac{\alpha }{M_{q}\pi }[ak_{\nu }+bM_{q}\gamma
_{\nu }+c\frac{k_{\nu }}{M_{q}}\hat{q}_{1}+d\hat{q}_{1}\gamma _{\nu }],
\label{VL} \\
\widehat{M}(p) &=&\frac{\alpha }{2M_{q}\pi }[-a+f\frac{1}{M_{q}}(\hat{p}%
+M_{q})](\hat{p}-M_{q})^{2},  \label{ML}
\end{eqnarray}%
with $\widetilde{t}%
=t/M_{q}^{2},~t=(p_{2}-q)^{2}-M_{q}^{2}=(p_{1}-q_{1})^{2}-M_{q}^{2},\quad
l_{t}=\ln (-\widetilde{t})$ and the coefficients are
\begin{eqnarray}
a &=&-\frac{1}{2(\widetilde{t}+1)}\left( 1-\frac{3\widetilde{t}+2}{%
\widetilde{t}+1}l_{t}\right) ,\quad b=-1-\ln \frac{\lambda }{M_{q}}-\frac{1}{%
2\widetilde{t}}R+\frac{\widetilde{t}+2}{4(\widetilde{t}+1)l_{t}}, \\
c &=&-\frac{1}{\widetilde{t}^{2}}R-\frac{\widetilde{t}+2}{2\widetilde{t}(%
\widetilde{t}+1)}+\frac{(\widetilde{t}+2)(2\widetilde{t}+1)}{2\widetilde{t}(%
\widetilde{t}+1)^{2}}l_{t},\quad d=-\frac{1}{2(\widetilde{t}+1)}l_{t},
\notag \\
f &=&\frac{1}{\widetilde{t}}\left( 1+2\ln \frac{\lambda }{M_{q}}+\frac{%
\widetilde{t}+2}{2(\widetilde{t}+1)}+\frac{\widetilde{t}^{2}-4\widetilde{t}-4%
}{2(\widetilde{t}+1)^{2}}l_{t}\right) ,  \notag \\
R &=&\frac{\pi ^{2}}{6}-\mathrm{Li}_{2}(\widetilde{t}+1).  \notag
\end{eqnarray}%
Note, that the fictive photon mass $\lambda ,$ introduced here, disappears
from the final answer.

\end{document}